**Ripple Texturing of Suspended Graphene Atomic Membranes**

Wenzhong Bao[1], Feng Miao[1], Zhen Chen[2], Hang Zhang[1], Wanyoung Jang[2], Chris Dames[2],

Chun Ning Lau[1*]

[1]Department of Physics and Astronomy, University of California, Riverside, CA 92521

[2]Department of Mechanical Engineering, University of California, Riverside, CA 92521

**Abstract**

Graphene is the nature's thinnest elastic membrane, with exceptional mechanical and electrical

properties. We report the direct observation and creation of one-dimensional (1D) and 2D

periodic ripples in suspended graphene sheets, using spontaneously and thermally induced longitudinal strains on patterned substrates, with control over their orientations and wavelengths.

We also provide the first measurement of graphene's thermal expansion coefficient, which is anomalously large and *negative*, $\sim -7 \times 10^{-6}$ K$^{-1}$ at 300K. Our work enables novel strain-based engineering of graphene devices.

* E-mail: lau@physics.ucr.edu

As the only known isolated atomic membrane[1-3], graphene offers a unique platform for understanding the fundamental mechanical properties of nano-materials, such as its extremely high elastic constants[4] and breaking strength[5], and the spontaneous formation of ripples in suspended graphene [6]. Unlike other thin films, the local mechanical distortions in graphene are expected to have a profound impact on its electrical properties, including creating effective magnetic fields[7, 8], modifying the local electronic potential[9, 10], breaking the charges' valley degeneracy[8], and inducing additional scattering[11, 12]. Such a close relationship between graphene's morphology and electrical properties, and its readiness to deform, could be explored to enable device tailoring based on manipulation of local strains[13], or to facilitate *selective* formation of $sp^3$ carbon-carbon bonds for bandgap creation[14]. However, even though corrugations have been inferred from finite-sized electron diffraction spots in suspended graphene sheets[6], ripples have neither been directly observed or controlled.

Here we report the direct observation and controlled creation of periodic ripples in suspended graphene sheets via spontaneously and thermally generated strains. We find that thin film mechanics, developed for continuum materials, continues to describe these atomically thin membranes. We manipulate the orientation, wavelength and/or amplitude of the ripples via controlled boundary conditions and the difference in the thermal expansion coefficients (TEC) between graphene and the substrate. This thermo-mechanical manipulation is especially effective because of graphene's anomalously large and negative TEC, which is measured to be ~5-6 times larger than that of bulk graphite in the basal plane. Finally, as a first step towards systematic investigation of ripple-related transport effects, we perform electrical measurements on these graphene devices. Our results indicate that small ripples do not introduce significant scattering, consistent with prior results[15, 16], and shed light on some of the unusual behaviors observed in ref. [16].

Graphene membranes, ranging from single layers (~0.3 nm) to ~ 16 nm in thickness, and ~0.5 to 20 μm in width, are suspended across pre-defined trenches on Si/SiO$_2$ substrates[17]. We examine their morphology under a scanning electron microscope (SEM) or an atomic force microscope (AFM). Strikingly, most of the graphene sheets are not flat, but spontaneously form nearly periodic ripples (Fig. 1a-c). Typically, the ripple crests are perpendicular to the edges of the trench (*y*-direction), although oblique ripples are also occasionally observed[17]. The out-of-plane displacement ζ of the ripples is well-described by a sinusoidal function,

$$\zeta = A\sin(2\pi y/\lambda). \qquad (1)$$

where *A* is the amplitude and *λ* the wavelength. We have imaged and measured more than 50 different membranes, with *A* ranging from 0.7 to 30 nm, and *λ* ranging from 370 nm to 5 *μ*m.

To understand the origin of these ripples, we note that for an elastic thin film, ripples described by Eq. (1) may be induced by either transverse compression in the *y* direction, or by longitudinal strain and/or shear in the *x*-direction[18]. From classical elasticity theory[19], we expect the clamped boundary conditions imposed by the banks of the trenches suppress lateral movement and induce local biaxial stress. For a thin film of thickness *t* with clamped boundaries at *x*=0 and *x*=*L*, the presence of a longitudinal tensile strain γ leads to[18]

$$\frac{\lambda^4}{(tL)^2} = \frac{4\pi^2}{3(1-\nu^2)\gamma} \quad \text{and} \quad \frac{A^2}{\nu tL} = \left[\frac{16\gamma}{3\pi^2(1-\nu^2)}\right]^{1/2} \qquad (2)$$

Here ν =0.165 is the Poisson ratio for graphite in the basal plane[20]. Combining both equations, we eliminate γ and obtain a relation with only experimentally accessible parameters:

$$\frac{A\lambda}{L} = \sqrt{\frac{8\nu}{3(1-\nu^2)}}t \qquad\qquad (3)$$

If instead the applied stress is dominated by in-plane shear, the equation takes a different prefactor[21]

$$\frac{A\lambda}{L} = \sqrt{\frac{8}{3(1+\nu)}}t \qquad\qquad (4)$$

Using values of *A, L, λ* and *t* as determined from AFM images, we plot *Aλ/L* vs. *t* for 51 devices that display periodic ripples, as shown by the data in Fig. 1d. Eqs. (3) and (4) are plotted as the lower and upper lines, respectively. Most of the data points fall on the lower solid line, indicating that the ripples are induced by pre-existing longitudinal strains in graphene. However, the 6 data points that fall above the upper line have a similar slope to the latter, suggesting the presence of shear in these devices.

Eqs. (2) – (4) are derived based on classical thin-film elasticity theory, and may not be valid *a priori* for atomically thin membranes. The inset of Fig. 1d displays *Aλ/L* vs. *t* for samples that are 1, 2 and 3 layers thick. Remarkably, the data points falls on a straight line, suggesting that Eq. (3) holds even for single atomic layer membranes.

The strains in these atomic membranes can also be readily obtained from Eq. (2). In Fig. 1e, we use the first equation to compute $\gamma$ for membranes with strain-induced ripples, and plot $\gamma$ vs *t*. For thicker films, $\gamma$ is relatively small, ~0.016% to 0.3%. In contrast, thinner films are more easily strained, and exhibit $\gamma$ up to 1.5%.

Building on our observation of periodic ripples in graphene membranes, we now show that these ripples can be controllably produced via simple thermal manipulation. The graphene membranes are annealed in a furnace in argon up to 700 K, and imaged again at room temperature. Surprisingly, almost all graphene membranes undergo one or both of the following dramatic changes in morphology – (1) the ripple geometry is significantly altered, with apparently larger amplitudes and longer wavelengths; (2) the graphene membrane buckles, typically sagging toward the substrate, or occasionally buckling upwards. In fact, the buckling can be quite dramatic: the central portions of several membranes settled on the bottom of the trenches *without breaking*[17].

To understand these observations, we perform *in situ* SEM imaging of our devices at different temperatures $T$, using a custom-built SEM stage with a built-in heater and a thermocouple. Fig. 2a-b show two image sequences for two different membranes. When $T$ is raised to 450-600K, the membranes are flat, and any pre-existing ripples completely disappear. However, upon cooling down to 300K, ripples invariably appear, usually with much larger amplitudes than any pre-annealing ones. The device in Fig. 2b also exhibits longitudinal buckling and sags into the trench. A movie of ripple formation in a single-layer graphene film during cooling is included online[17].

The above observations suggest that, after thermal annealing, a graphene sheet experiences biaxial compression[22]; the different behaviors (rippling *vs.* buckling) arise from the different boundary conditions in the two directions. This process may be understood in terms of graphene's thermal *contraction*, and the competition between three forces: (1). $F_{pin}$, the substrate-pinning force that prevents the graphene membrane from sliding; (2) $F_b$, the bending/buckling critical compression force, which is generally $<< F_{pin}$; and (3). $F_{stretch}$, the elastic restoring force under tension. Fig. 2c shows a schematic of the process. When $T$ increases, the substrate and the trench width expand biaxially, while graphene *contracts*; this differential in TEC places the membrane in biaxial tension. Once $F_{stretch} > F_{pin}$, the taut membrane slides *irreversibly* over the substrate into the trench, hence "erasing" any pre-existing ripples.

Conversely, the cooling process applies compressive stress; since $F_b < F_{pin}$, the ends of the graphene remain pinned to the banks of the trench, resulting in transverse ($y$) ripples and/or longitudinal ($x$) buckling.

Such interplay between the thermal expansion of the substrate and the membrane suggest a simple way to control both the amplitude (or, if desired, the wavelength) and orientation of the ripples. Since the membranes buckle readily under compression, the transverse compressive strain[18] is $\Delta \sim \sqrt{1 + \frac{\zeta^2}{\lambda^2}} - 1$. Hence $A$ and $\lambda$ of the post-annealed wrinkles are related, $A \sim \lambda\sqrt{\Delta}$ for $A << \lambda$. Since $\Delta$ arises from the difference in TEC between the substrate and graphene, we expect $\Delta$ to scale with $\Theta_{max}$, the maximum annealing temperature rise above ambient. Fig. 3a plots $A$ vs $\lambda\sqrt{\Theta_{max}}$ for 6 different devices, each thermally cycled to several different temperatures. Indeed, the data points fall approximately on a straight line. Thus, for a given set of boundary conditions, the ripple's $A$ and $\lambda$ can be controlled by $\Theta_{max}$.

To control the orientation of the ripples, we note that the ripple patterns are determined by the substrate-imposed boundary conditions (*e.g.* buckling *vs.* rippling in $x$ and $y$-directions, respectively). This is similar to that in metallic thin films on elastomeric substrates that were patterned with relief structures[23]. In both experiments, ripples patterns, in which crests are aligned perpendicular to the step-like structures on the substrate, arise from the redistribution of compressive stresses due to the TEC differential between the substrate and thin film. Hence, as the first step towards controlled creation of 2D ripples, we pattern openings of different shapes on the substrates. Graphene membranes are suspended over these openings; annealing in temperature up to 700K yields striking patterns of 2D ripples, with the crests perpendicular to the edges of the opening (Fig. 3b). Such 1D or 2D ripple patterns may be desirable for novel devices such as in-plane electronic superlattices[24, 25]. In the long term, just as the creation of complex patterns was demonstrated in ref.[23], simple thermal manipulation, coupled with pre-patterned relief structures on substrates, can be used to engineer graphene's local morphology and alter its electronic properties. Such processes are also compatible with large-scale device applications.

We now explore the interplay between the ripple mechanics and graphene's thermodynamical and electrical properties. Notably, our thermo-mechanical control of the amplitude and orientation of the ripples proves to be exceedingly effective, due to graphene's *negative* TEC that accentuates the TEC-difference between the substrate and graphene. Indeed, a negative TEC is expected as a consequence of graphene's two-dimensionality, in which the energies of out-of-plane (bending) phonon modes is lower for smaller lattice parameters[26]. However, although graphene's TEC has been calculated[27-29], no experimental measurement has been reported to date.

Our experiment readily enables measurement of the TEC $\alpha(T)$. To this end, we anneal a single-layer graphene sheet in a furnace up to 700K to create a sagging membrane. This device is then inserted into the SEM chamber, and heated up to ~450 K, at which the membrane is taut across the trench. The heater is then turned off to allow the membrane to cool to 300K. At a given temperature $T$, we take an image of the sagging membrane, and compute the ratio $l(T)=L_g(T)/L_t(T)$, where $L_g$ is the length of graphene membrane as measured along the arc, and $L_t$ is the length of the trench measured along the chord of the arc (Fig. 4a inset). Both quantities are measured independently for every image to minimize errors induced by, *e.g.,* slight variations in the imaging conditions. In Fig. 4a we plot $l(T)$ for a single layer graphene sheet. The slope of the graph can be approximated by $b = \dfrac{dl}{dT} \approx \alpha - \alpha_{Si}$, where $\alpha_{Si}$ is silicon's TEC. To obtain $b$, the data points are approximated by an analytical function, which is then differentiated. Two different fitting functions are used to illustrate the error range of this procedure. Using values of $\alpha_{Si}(T)$ from ref. [30], we can determine $\alpha(T)$ for graphene (lower panel of Fig. 4a, blue lines). Our measurements indicate that at 300K, $\alpha \sim -7 \times 10^{-6}$ K$^{-1}$. This agrees with our expectation that $\alpha$ is much larger than that of graphite in the basal plane, $\sim -1 \times 10^{-6}$ K$^{-1}$. However, the measured $\alpha$ is roughly twice the value predicted from theoretical calculation, and approaches zero more quickly than expected[27]. Part of the discrepancy may be attributed to our unconventional method to determine $\alpha$; nevertheless, this first quantitative measurement of graphene's TEC provides important insight into graphene's unique thermal properties.

Finally, we note that our ability to controllably create ripples in graphene opens a door for systematic investigation of ripples' effects on graphene's electrical properties, and strain-based engineering of graphene devices. For instance, despite intense theoretical interest[8, 31-34], the nature and *magnitude* of ripples' effects on electron scattering processes have not been experimentally studied, and remain controversial to date. As an initial demonstration of such studies, we fabricate suspended graphene devices with two electrodes on each side of the trench (Fig. 4b), allowing measurements of both suspended and substrate-supported portions of the same graphene sheet. Fig. 4c shows the 2-probe conductance $G$ of a single layer graphene device with small random ripples as a function of the charge density $n$. Comparing with the substrate-supported part, the suspended part of graphene has much higher mobility, with a sharper Dirac point that is closer to $n=0$, indicating smaller density of charged impurities. Thus, our measurements confirm that, despite the presence of small random ripples, the device mobility is substantially enhanced by the elimination of the substrate. This is consistent with prior results from high-mobility suspended graphene devices[15, 16], which, in light of our results, are also likely to contain ripples. Our observation of the temperature-dependent morphology of suspended graphene sheets may also have implications on understanding results reported in ref. [16], in which suspended graphene devices display unusual $G(T)$ behaviors with large sample-to-sample variations.

We thank Marc Bockrath and Shan-Wen Tsai for discussions, and Bophan Chim, Dong Yan and Zeng Zhao for assistance with SEM imaging. This research is supported in part by NSF/CAREER DMR/0748910, NSF/CBET 0756359 and ONR/DMEA H94003-07-2-0703.

**Figure 1.** (a, b) Data from two different suspended graphene membranes suspended across trenches. Upper panels: AFM topographical images. Lower panels: line traces taken along the dotted lines. Note the different amplitudes and wavelengths of the devices. (c) SEM image of a bi-layer suspended membrane. (d) $A\lambda/L$ vs $t$ for 51 membrane devices. The lower and upper lines are calculated using $\nu$=0.165 and Eqs. (3) and (4), respectively. Inset: $A\lambda/L$ vs $t$ for single(red)-, bi(green)- and triple(blue)- layer devices. The number of layers is inferred from color contrast in optical microscope, though only measured thickness is used. (e) Strain in suspended devices, calculated using Eq. (2).

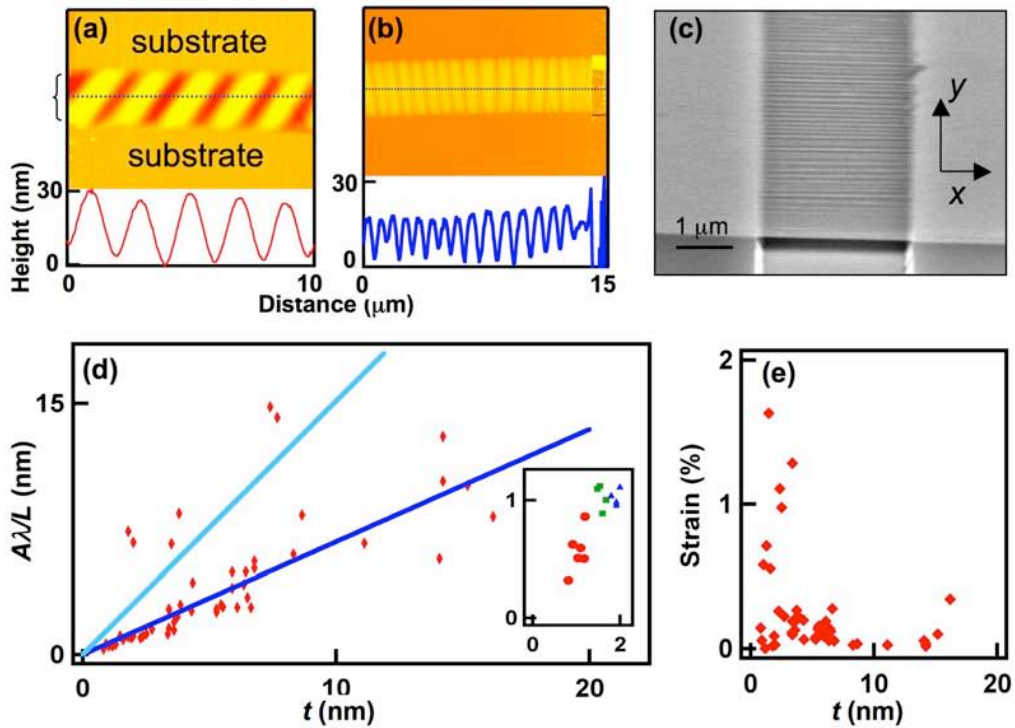

**Figure 2.** Graphene membranes before (left panels), during (middle) and after (right) annealing. (a, b) *In situ* SEM images of two devices. Bottom panels of (b) are higher magnification images of the edge of the graphene membrane, which sags into the trench after annealing. (c) Schematic of buckling of a graphene membrane. The arrows indicate the contraction/expansion of the substrate and graphene.

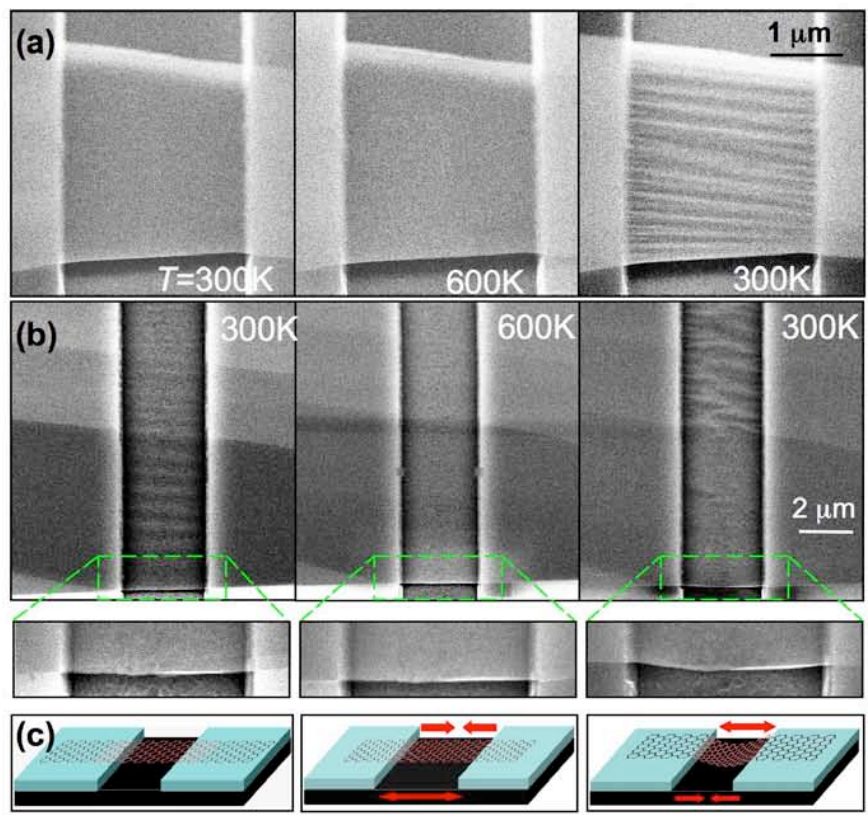

**Figure 3.** (a) *A vs. λΘ$_{max}$$^{1/2}$* for post-annealed devices. (b) Formation of periodic 2D ripples in graphene membranes suspended over openings of various shapes. Scale bars: 1 μm.

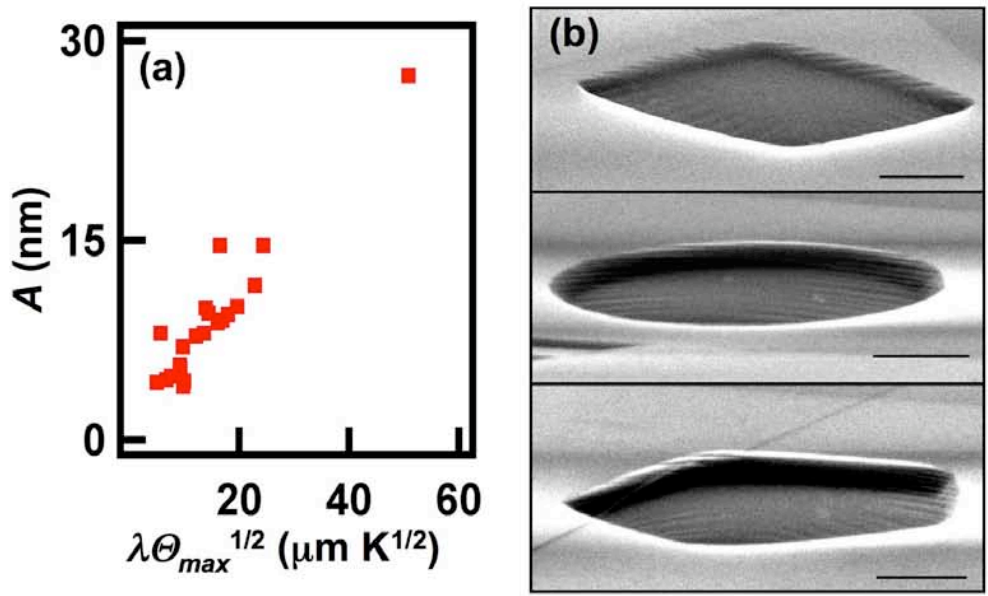

**Figure 4.** (a) Upper Panel: *l(T)* for a single layer graphene membrane. The solid line is a 4th-order polynomial fit to the data points, and the dotted line is an exponential function fit. Inset: an SEM image of a sagging few-layer graphene sheet. Lower Panel: Slope (red) and TEC (blue) of a single-layer graphene membrane. The solid and dotted lines correspond to results obtained using the polynomial and exponential functions, respectively. Silicon's TEC is obtained from ref. [30] and plotted as the green dotted line. (b) SEM image of a bi-layer graphene device with small random ripples. (c) $G(n)$ at 1.5K for suspended (red) and substrate-supported (blue) part of a single layer graphene device with small random ripples ($A \sim 5\text{-}15$ nm, $\lambda \sim 0.6 - 2$ μm). $n/V_g \sim 2.3\text{x}10^{10}$ and $7.2$ x$10^{10}$ cm$^{-2}$V$^{-1}$ for suspended and supported graphene sheets, respectively. Scale bars in all panels: 1 μm.

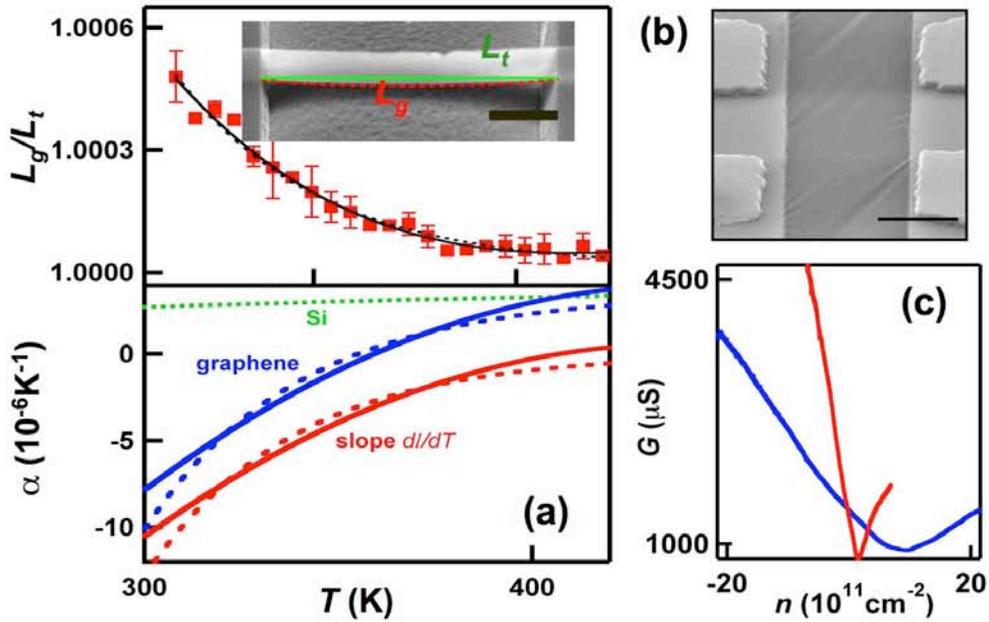